\pdfoutput=1

\documentclass[prl,superscriptaddress,twocolumn,showpacs,showkeys,amsmath,amsthm,amsfonts,amssymb,nofootinbib]{revtex4} 
\usepackage{graphicx}

\newcommand{\eqsp}{=}  
\newcommand{\tosp}{\to}
\newcommand{\llsp}{\ll}
\newcommand{\ltsp}{<}

\begin{document}

\title{Non-asymptotic thermodynamic ensembles}

\author{Robert K. Niven}
\affiliation{School of Aerospace, Civil and Mechanical Engineering, The University of New South Wales at ADFA, Northcott Drive, Canberra, ACT, 2600, Australia.}
\email{r.niven@adfa.edu.au}

\date{5 March 2009}

\begin{abstract}
Boltzmann's principle is used to select the ``most probable'' realization (macrostate) of an isolated or closed thermodynamic system, containing a small number of particles ($N \llsp \infty$), for both classical and quantum statistics.  The inferred probability distributions provide the means to define intensive variables and construct thermodynamic relationships for small microcanonical systems, which do not satisfy the thermodynamic limit. This is of critical importance to nanoscience and quantum technology. 
\end{abstract}



\maketitle   


\section{Introduction} 

In 1877, Boltzmann \cite{Boltzmann_1877} discovered the combinatorial basis of entropy, usually expressed as \cite{Planck_1901}:
\begin{equation}
S_{total} =k \ln \mathbb{W},
\label{eq:Boltzmann1}
\end{equation}
where $S_{total}$ is the total thermodynamic entropy of a system, $k$ is the Boltzmann constant and $\mathbb{W}$ the statistical weight, i.e.\ the number of ways in which a given realization (macrostate) of the system can occur, as defined by the 
number of particles $n_i$ in each category $i=1,...,s$
of the system. Eq.\ \eqref{eq:Boltzmann1} can be rewritten in the dimensionless form:
\begin{equation}
H = \frac{S_{total}}{kN} = \frac{1}{N} \ln \mathbb{W},
\label{eq:Boltzmann2}
\end{equation}
where $H$ is the dimensionless entropy per particle and $N$ is the (actual) number of particles \cite{Niven_MaxEnt07}. 
Eq.\ \eqref{eq:Boltzmann2} extends naturally to the probabilistic definition \cite{Niven_MaxEnt07, Vincze_1972, Grendar_G_2001}:
\begin{equation}
-D=\frac{1}{N} \ln \mathbb{P},
\label{eq:Boltzmann3}
\end{equation}
where $D$ is the divergence or cross-entropy of the system, per unit particle, and $\mathbb{P} = P(\{n_i\}|\{q_i\},N)$ is the probability of occurrence of a given realization, subject to $N$ and the source distributions 
(``prior probabilities'') $q_i$ of each category.
Maximisation of $H$ (MaxEnt) or minimisation of $D$ (MinXEnt), subject to its constraints, therefore selects the realization of highest weight $\mathbb{W}$ or probability $\mathbb{P}$, a technique which can be termed the {\it maximum probability principle} (MaxProb) \cite{Boltzmann_1877, Planck_1901, Vincze_1972, Grendar_G_2001, Niven_MaxEnt07}.  The inferred distribution is then used to represent the system. Typically, $\mathbb{W}$ or $\mathbb{P}$ are considered to follow the multinomial weight or distribution:
\begin{eqnarray}
\mathbb{W}_{mult} = N! \prod\limits_{i = 1}^s \frac{1 }{n_i !}, 
\label{eq:multinomialW}
\\
\mathbb{P}_{mult} = N! \prod\limits_{i = 1}^s \frac{q_i ^{n_i } } {n_i !} ,
\label{eq:multinomialP}
\end{eqnarray}
obtained either by a ``frequentist'' model of the system, or by a Bayesian inferential method involving a weighted sum of all possible models \cite{Blower_2007, Niven_MaxEnt07}. In these cases,
$H$ and $D$ converge respectively in the asymptotic limit $N \to \infty$ 
(by the Sanov theorem \cite{Sanov_1957})
to the Shannon \cite{Shannon_1948} entropy or Kullback-Leibler \cite{Kullback_L_1951} cross-entropy functions: 
\begin{eqnarray}
H_{Sh}  =  \lim_{N \to \infty} \frac{1}{N} \ln \mathbb{W}_{mult} =  - \sum\limits_{i = 1}^s {p_i \ln p_i } 
\label{eq:Shannon}
\\
-D_{KL}  =  \lim_{N \to \infty} \frac{1}{N} \ln \mathbb{P}_{mult} = - \sum\limits_{i = 1}^s {p_i \ln \frac{ p_i }{q_i}}, 
\label{eq:KL}
\end{eqnarray}
where $p_i=n_i/N$ is the 
frequency or probability of occupancy of the $i$th category.
This provides a probabilistic justification for these functions (based on $\mathbb{W}$ or $\mathbb{P}$), independent of the standard axiomatic derivations given in information theory 
\cite{Shannon_1948, Shore_J_1980}. The cross-entropy \eqref{eq:Boltzmann3} or \eqref{eq:KL} contains the source distributions $q_i$, and is thus more general than the entropy \eqref{eq:Boltzmann2} or \eqref{eq:Shannon}; 
in thermodynamics, this is often handled by taking $q_i = g_i/G$, where $g_i$ is the number of distinguishable subcategories in category $i$ (its degeneracy) and $G=\sum\nolimits_{i=1}^s g_i$.

It is of interest to consider systems of small numbers of particles $N \ll \infty$, of critical importance in nanoscience, biochemistry and  quantum technology. Such systems will not satisfy the ``thermodynamic limit'', in which both $N$ and the system volume tend to infinity whilst the particle density remains constant \cite{Hill_1956, Hill_1963, Buchdahl_1975}. For such a system, is it possible to infer a representative distribution of particles amongst its categories?  Inference using \eqref{eq:Shannon} or \eqref{eq:KL} is not possible, since these require the asymptotic limit $N \tosp \infty$.  From the above discussion, it is clear that inference must proceed by the MaxProb principle - involving extremisation of the entropy or cross-entropy defined by \eqref{eq:Boltzmann2}-\eqref{eq:Boltzmann3} - since this invokes a simple probabilistic proposition, which is independent of (indeed, it defines) thermodynamic concepts. 
Although the inferred distribution will not be as dominant as in the asymptotic case - i.e.\ the ``most probable'' will not be the ``only observable'' distribution - this should not deter us from conducting this analysis, nor prevent us from enlarging the body of thermodynamics to explore the effect of $N$.

The aim of this study is to initiate 
a new non-asymptotic formulation of thermodynamics for small $N$ systems, based exclusively on the MaxProb principle. 
This differs markedly from ensemble-based formulations for small systems, such as by Hill \cite{Hill_1963}. The new approach has the advantages of a strong foundation in probability theory, and being {\it directly} applicable to individual systems of particles rather than ensembles of systems, both within and beyond the domain of thermodynamics.
In turn, we examine an isolated (\S2) and a closed (energy-diffusive) (\S3) small system, subject to an energy constraint. The analyses are related, asymptotically, to the microcanonical ensemble and certain features of the canonical ensemble. Conclusions are drawn concerning the system temperature and zeroth law of thermodynamics. In \S4, the analysis is extended to quantum systems governed by Bose-Einstein (BE) or Fermi-Dirac (FD) statistics. The results have important implications for the thermodynamics of small systems.

\section{Isolated Systems} 
\begin{figure}[t]
\begin{center}
\setlength{\unitlength}{0.6pt}
  \begin{picture}(280,370)
   \put(25,190){\includegraphics[width=45mm]{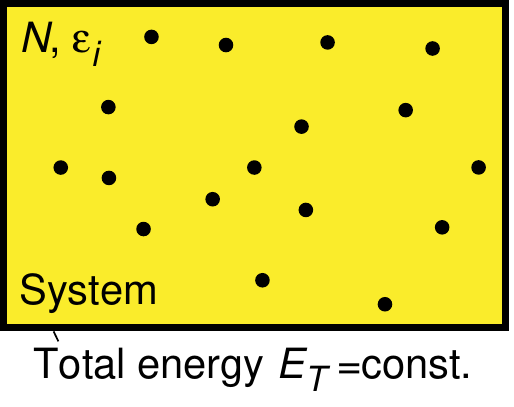} }
   \put(0,190){\small (a)}
   \put(25,0){\includegraphics[width=45mm]{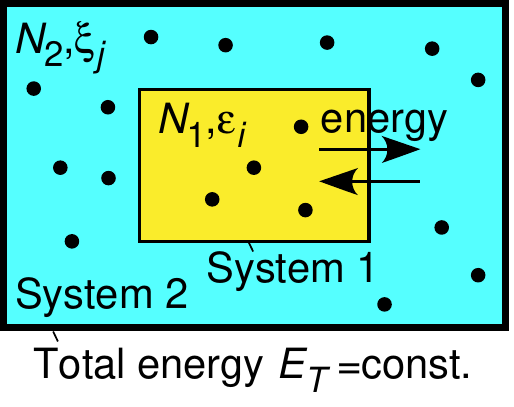} }
   \put(0,0){\small (b)}
  \end{picture}
\end{center}
\caption{Non-asymptotic analyses of (a) an isolated system, and (b) a closed (energy-diffusive) system.}
\label{fig:systems}
\end{figure}

We first consider a single, isolated system of $N \ltsp \infty$ particles enclosed by a particle- and energy-impermable wall, of constant total energy $E_T$, as shown in Figure \ref{fig:systems}a.  The particles are distributed amongst energy levels $\epsilon_i, i=1,...,s$, of source probabilities $q_i$. 
In the first instance,
the filling of particles in levels is assumed to follow classical multinomial statistics \eqref{eq:multinomialP} 
(n.b. quantum statistics are considered in \S\ref{Quantum}).
The system is therefore the non-asymptotic form of the microcanonical ensemble.  From the MaxProb principle \eqref{eq:Boltzmann3}, the cross-entropy is:
\begin{equation}
-D=\frac{1}{N} \ln \mathbb{P} = \frac{1}{N} \Bigr\{ \ln N! + \sum\limits_{i = 1}^s \bigr[ n_i \ln q_i  - \ln n_i ! \bigr] \Bigr\}.
\label{eq:Dx}
\end{equation}
which is subject to the constraints:
{\begin{eqnarray}
\sum\limits_{i = 1}^s n_i = N,
\label{C0}
\\
\sum\limits_{i = 1}^s n_i \epsilon_i = E_T = \langle E \rangle N. 
\label{C1}
\end{eqnarray}
where $\langle E \rangle$ is the mean energy per particle}.  Applying the calculus of variations to \eqref{eq:Dx}-\eqref{C1} gives the Lagrangian:
\begin{eqnarray}
L= \frac{1}{N} \sum\limits_{i = 1}^s \Bigr[ \frac{n_i}{N} \ln N!  +  n_i \ln q_i  - \ln n_i ! \Bigr]  - \kappa_0 \Bigl\{ \sum\limits_{i = 1}^s n_i - N \Bigr\} 
\\ \nonumber \qquad 
- \kappa_1 \Bigl\{ \sum\limits_{i = 1}^s n_i \epsilon_i - E_T \Bigr\}  
\label{eq:L}
\end{eqnarray}
where $\kappa_0$ and $\kappa_1$ are Lagrangian multipliers associated with constraints \eqref{C0}-\eqref{C1}, and the leading $\ln N!$ term is brought inside the sum using \eqref{C0}.  The extremum of \eqref{eq:L}, defined by $\delta L=0$, gives $\partial L / \partial n_i = 0, \forall i$ for constant $N$; this yields the most probable non-asymptotic distribution for the system:
\begin{equation}
p_i^{\#} = \frac{n_i^{\#}}{N} = \frac{1}{N} \, \Lambda^{-1} \Bigl[ \frac{1}{N} \ln N! + \ln q_i - \lambda_0 - \lambda_1 \epsilon_i \Bigr]  
\label{eq:phash}
\end{equation}
where $\lambda_j=\kappa_j N$ for $j \in \{0,1\}$ are modified Lagrangian multipliers, whilst $\Lambda^{-1}(y)= \psi^{-1}(y-1)$ is the upper inverse of the function $\Lambda(x)=\psi(x+1)$, 
wherein $\psi(x)$ is the digamma function. Note \eqref{eq:phash} is also obtained if one extremises $\ln \mathbb{P}$ instead of $D$.  There is no factorisable partition function, hence \eqref{eq:phash} must be solved simultaneously with both constraints \eqref{C0}-\eqref{C1}. 

In the asymptotic limit $N \to \infty$ (e.g.\ Sanov's \cite{Sanov_1957} theorem), extremisation of the Kullback-Leibler function \eqref{eq:KL} subject to \eqref{C0}-\eqref{C1} yields the Boltzmann distribution:
{\begin{equation}
p_i^{*} = \frac{n_i^{*}}{N} = q_i \exp ( - \lambda_0^{KL} - \lambda_1 \epsilon_i ) = \frac{ q_i \exp (- \lambda_1 \epsilon_i ) } {\sum\nolimits_{i=1}^s  q_i \exp (- \lambda_1 \epsilon_i ) }
\label{eq:pstar}
\end{equation}
where $\lambda_0^{KL}=\lambda_0+1$}.  As a corollary, \eqref{eq:phash} must converge to \eqref{eq:pstar} as $N \tosp \infty$. 

\begin{figure*}[t]
\begin{center}
\setlength{\unitlength}{0.6pt}
  \begin{picture}(675,550)
   \put(450,0){\includegraphics[height=50mm]{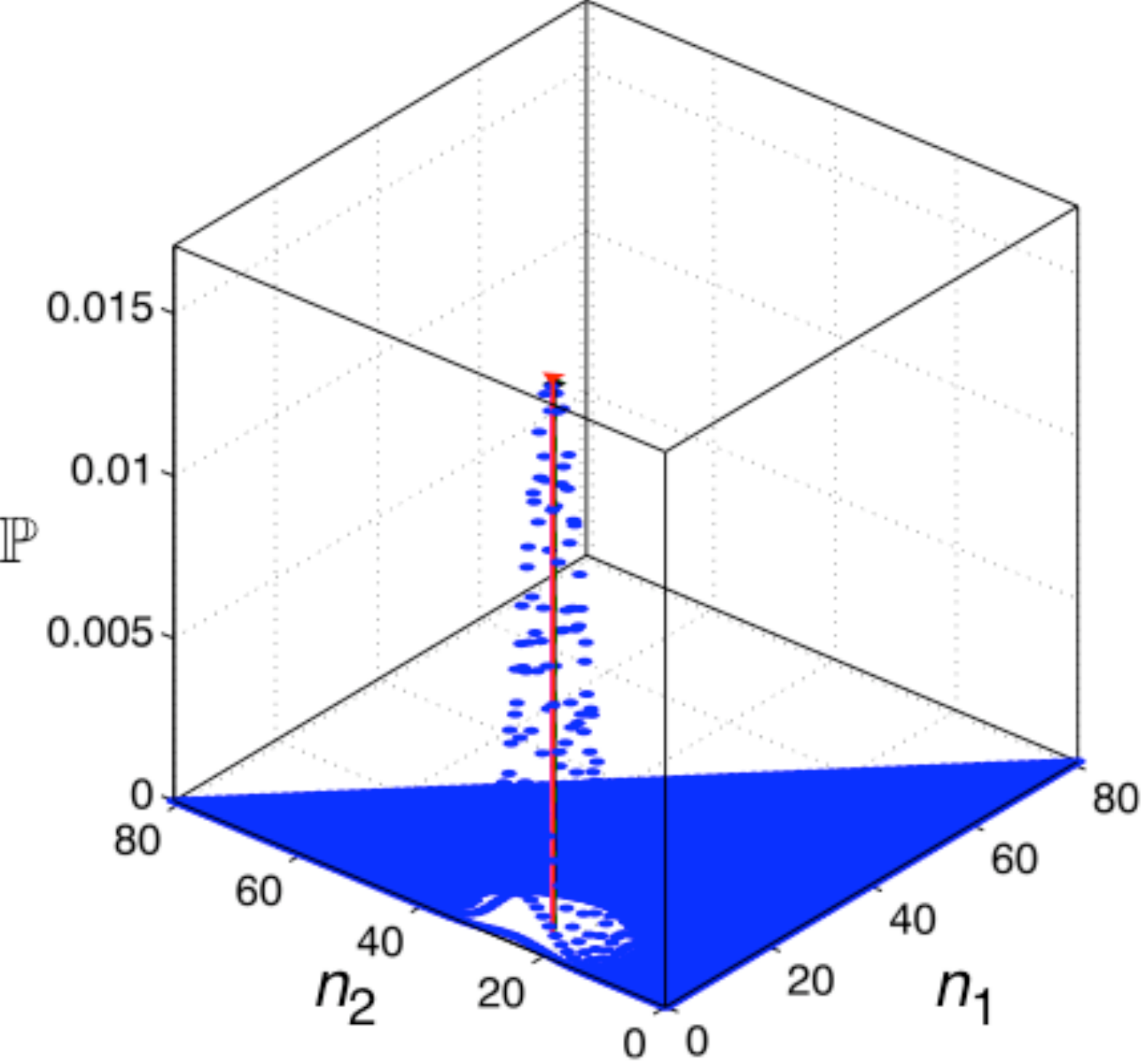} }
   \put(440,0){\small (d)}
   \put(80,0){\includegraphics[height=50mm]{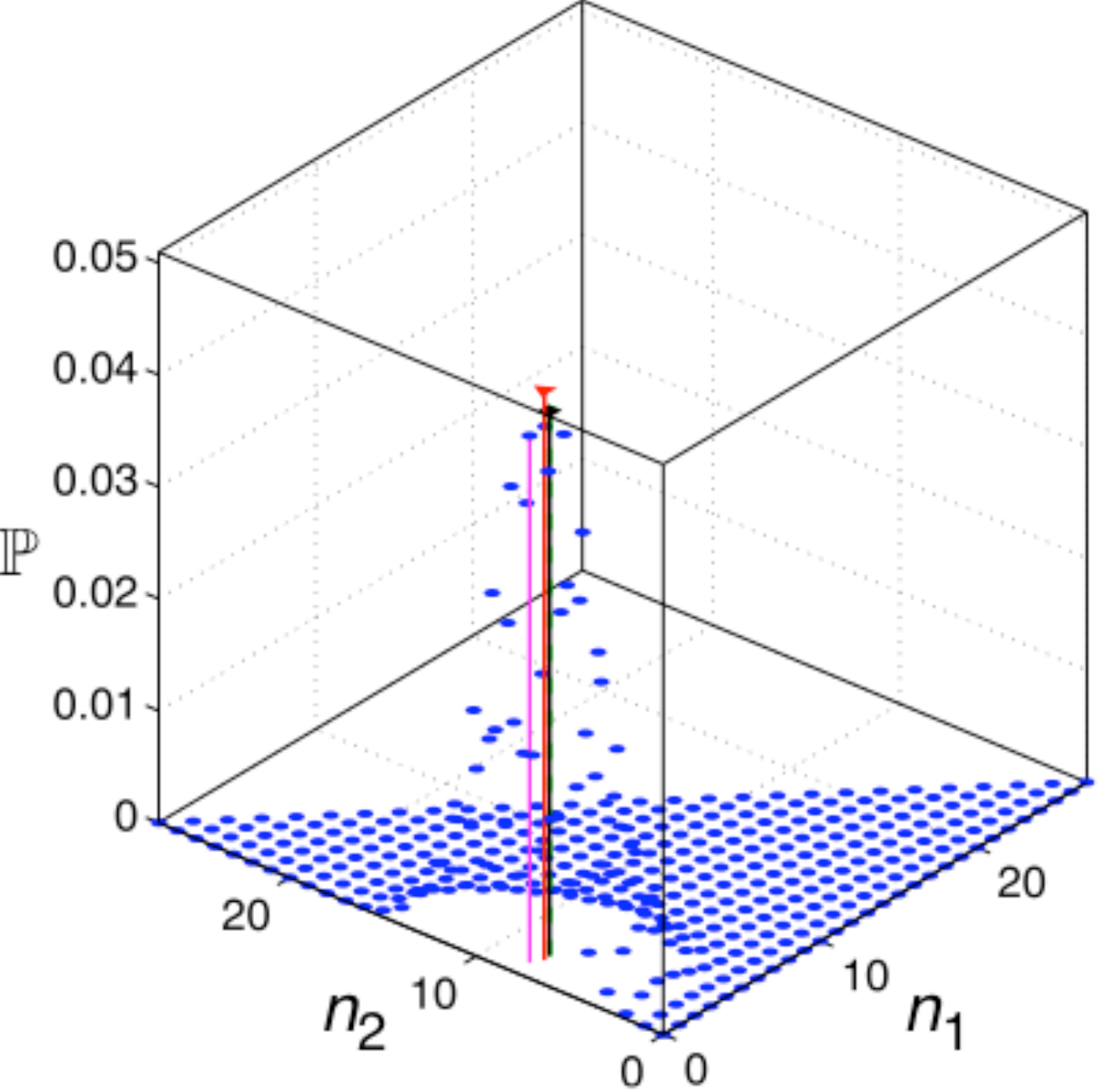} }
   \put(70,0){\small (c)}
   \put(450,290){\includegraphics[height=50mm]{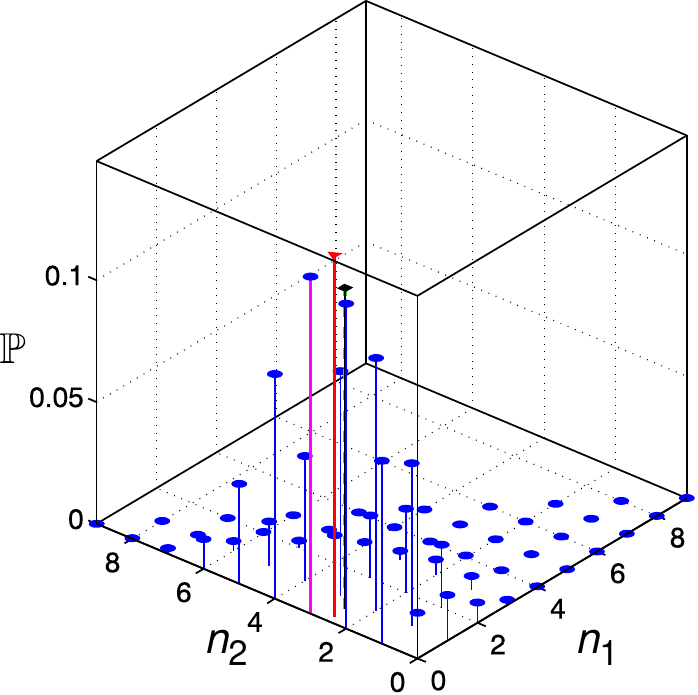} }
   \put(440,290){\small (b)}
   \put(80,290){\includegraphics[height=50mm]{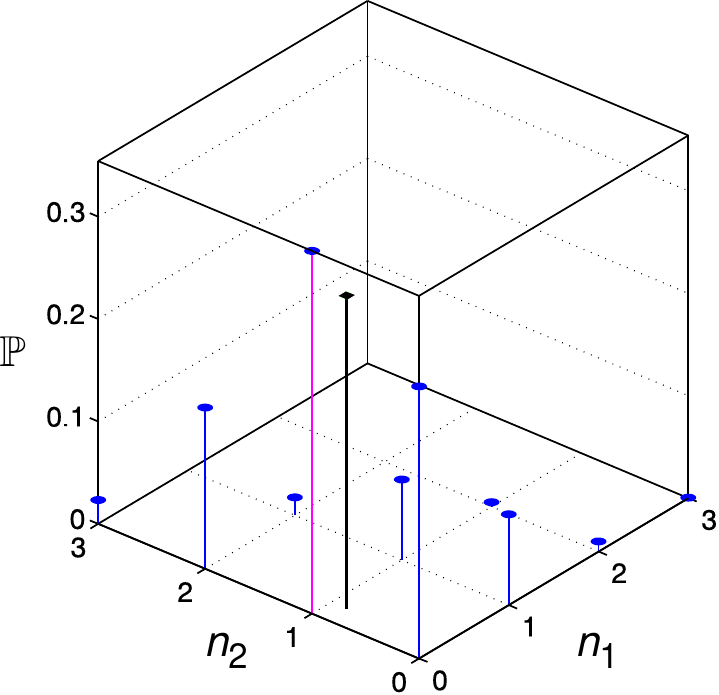} }
   \put(70,290){\small (a)}
      \put(320,440){\includegraphics[width=30mm]{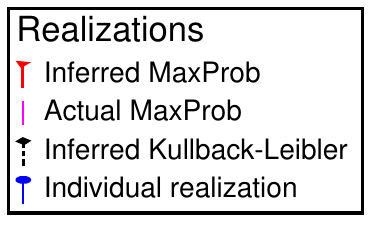} }
  \end{picture}
\end{center}
\caption{Example multinomial system subject to the natural constraint (see text) for (a) $N \eqsp 3$, (b) $N \eqsp 9$, (c) $N \eqsp 27$ and (d) $N \eqsp 100$, showing probability $\mathbb{P}$ of each realization, and inferred non-asymptotic \eqref{eq:phash} and asymptotic \eqref{eq:pstar} distributions.}
\label{fig:multinomial}
\end{figure*}
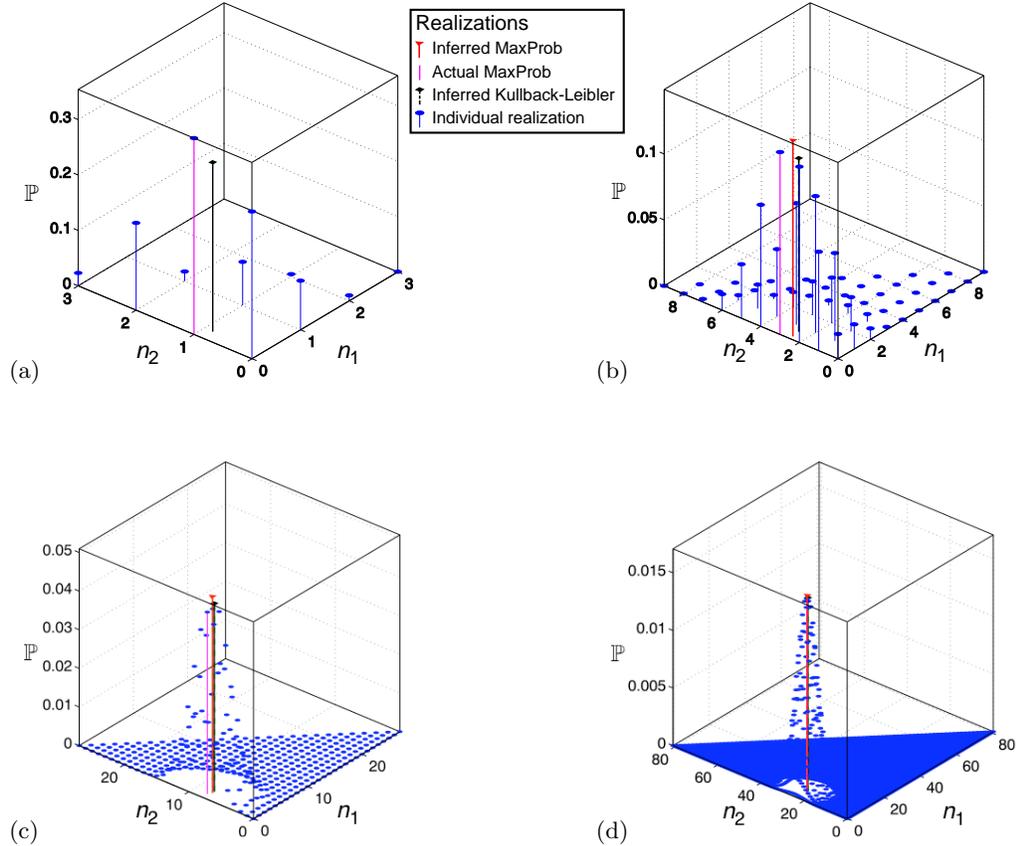

The character of a non-asymptotic system can be illustrated by two examples. Firstly, consider a system of $N$ particles with three energy levels of degeneracies ${\bf g}=[1,4,9]$, subject only to the natural constraint \eqref{C0}. The probabilities $\mathbb{P}$ of each realization $[n_1, n_2, n_3]$, calculated by \eqref{eq:multinomialP} for different $N$ {using a combinatorial search scheme, are illustrated by ``mortarboard plots''} against $n_1$ and $n_2$ in Figures \ref{fig:multinomial}a-d.  The inferred non-asymptotic \eqref{eq:phash} and asymptotic \eqref{eq:pstar} distributions are also shown.  As evident, for large $N$ (Figure \ref{fig:multinomial}d), the set of realizations is highly concentrated around the asymptotic distribution, and the non-asymptotic distribution converges to this peak. However, for small $N$ (Figures \ref{fig:multinomial}a-b), the realizations are more infrequent, with the system less dominated by its most probable realization. {At low $N$}, the inferred non-asymptotic distribution also becomes distinct from the asymptotic; moreover, due to the quantisation of levels, it may not coincide with the true most probable realization (the highest peak), but may {lie within} some neighbourhood of it.  {Even in this distinctly discrete case, the predicted distribution \eqref{eq:phash} - being the predicted most probable distribution of the system - can be used for further inference about the system.}

\begin{figure}[h]
\begin{center}
\setlength{\unitlength}{0.6pt}
  \begin{picture}(380,540)
   \put(20,0){\includegraphics[height=48mm]{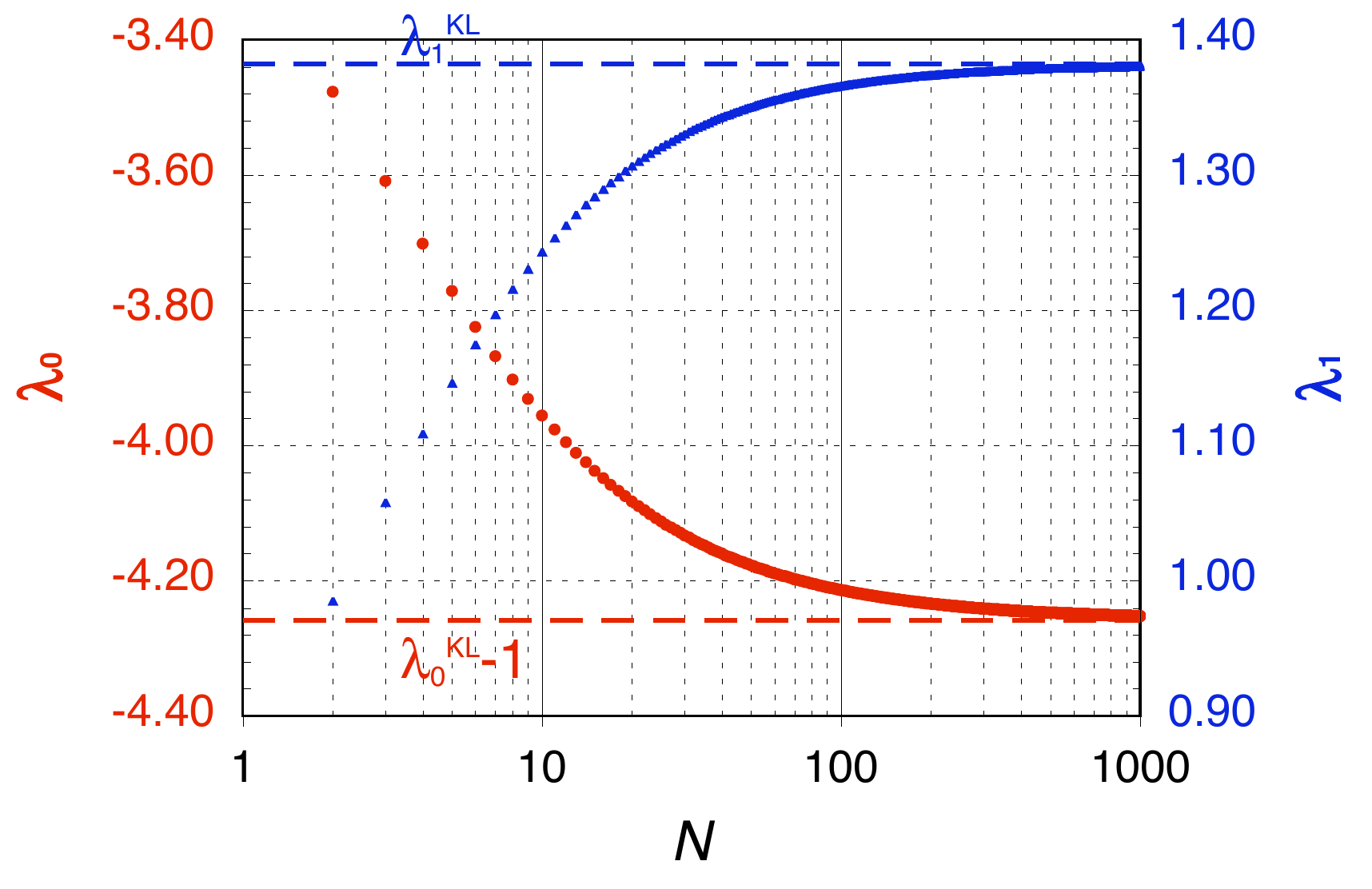} }
   \put(10,0){\small (b)}
   \put(40,280){\includegraphics[height=50mm]{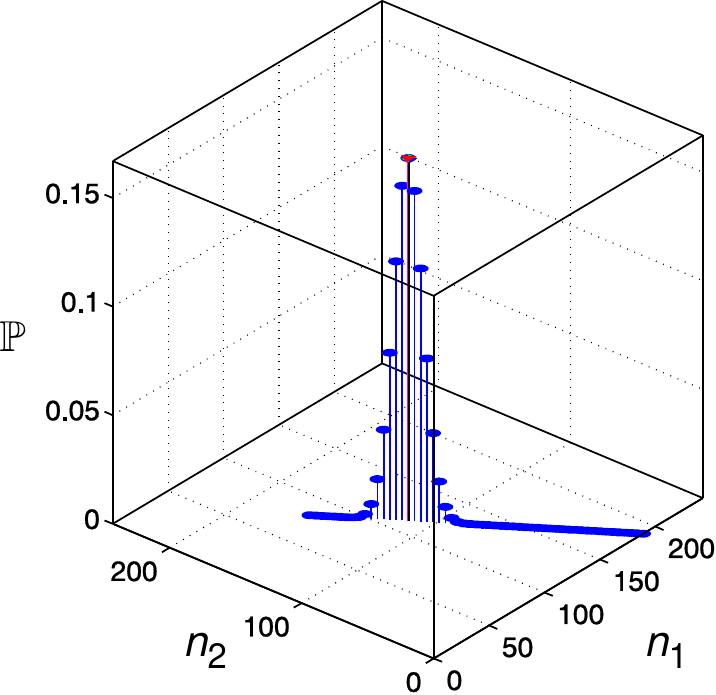} }
   \put(10,280){\small (a)}
  \end{picture}
  \end{center}
\caption{Example multinomial system subject to the natural and energy constraints with $\langle E \rangle=\frac{5}{3}$ (see text): (a) possible realizations for $N=243$; and (b) variation of $\lambda_0$ and $\lambda_1$ with $N$.}
\label{fig:mult2}
\end{figure}

{Secondly, consider the above multinomial system subject also to the energy constraint \eqref{C1} $\langle E \rangle = \frac{5}{3}$ or $2$, for which plots of each realization for various $N$ are given respectively in Figures S2-S3 in the Supplementary Data. The energy levels are taken as ${\mathbf \epsilon}=[1,2,4]$. One example is shown in Figure \ref{fig:mult2}a. Since the energy constraint excludes many realizations, it is necessary to renormalise the probabilities calculated using \eqref{eq:multinomialP}.  Further plots, to show the effect of $\langle E \rangle$ at $N=729$, are given in Figure S4; note that in this system, an asymptotic uniform distribution ($\lambda_1=0$ or infinite temperature) corresponds to $\langle E \rangle = 3.214$. As evident, imposing the energy constraint causes a dramatic reduction (``pruning'') in the number of available realizations. This substantially expands the non-asymptotic domain (to much higher $N \lessapprox 1000$ to $5000$), widening the separation between the inferred asymptotic and non-asymptotic distributions, and increasing the effect of quantisation.  Plots of the effect of $N$ and $\langle E \rangle$ on the non-asymptotic multipliers $\lambda_0$ and $\lambda_1$ are also given in Figure S5, of which one case is shown in Figure \ref{fig:mult2}b.  In all cases, the Massieu function $\lambda_0$ converges to its known asymptotic value $\lambda_0^{KL}-1$ as $N \to \infty$. Similarly, the inverse temperature $\lambda_1 =1/kT$ converges to a constant value (related to $\langle E \rangle$) as $N \to \infty$. Depending on the value of $\langle E \rangle$, the multipliers could converge in either direction, i.e.\ from positions of lower or} {higher free energy and/or from lower or higher temperature. Several peaked convergence curves in $\lambda_0$ are also observed.}

\section{Closed (Energy-Diffusive) Systems} 
Now consider two systems in contact, as shown in Figure \ref{fig:systems}b, in which System 1 contains $N_1$ particles with energy levels $\epsilon_i, i=1,...,s$ of source distributions $q_i$, whilst System 2 contains $N_2$ particles with energy levels $\xi_j, j=1,...,t$ of source distributions $r_j$; this could represent two systems in contact, or a single system in contact with an energy bath. The double system is enclosed by an impermable wall containing total energy $E_T$; within this, the systems make contact via a particle-impermeable but energy-permeable wall.  For multinomial statistics, the probabilities of realization $\{n_i\}$ in System 1 and realization $\{m_j\}$ in System 2 are respectively:
\begin{eqnarray}
\mathbb{P}_{1} = N_1! \prod\limits_{i = 1}^s \frac{q_i ^{n_i } } {n_i !},
\label{eq:P_sys1}
\\
\mathbb{P}_{2} = N_2! \prod\limits_{j = 1}^t \frac{r_j ^{m_j } } {m_j !} 
\label{eq:P_sys2}
\end{eqnarray}
If there is no correlation between occupancies in Systems 1 and 2, $\mathbb{P}_{1}$ and $\mathbb{P}_{2}$ are independent, whence the joint probability of the double realization $\{ \{n_i\}, \{m_j\}\}$ is:
\begin{eqnarray}
\mathbb{P}_{12} = \mathbb{P}_{1} \mathbb{P}_{2} = \Bigl\{ N_1! \prod\limits_{i = 1}^s \frac{q_i ^{n_i } } {n_i !} \Bigr\} \Bigl\{ N_2! \prod\limits_{j = 1}^t \frac{r_j ^{m_j } } {m_j !} \Bigr\}
\label{eq:P_sys12}
\end{eqnarray}
Note that in this case, it is necessary to analyse the double system using raw probabilistic principles, to ensure consistency of reasoning \cite{Niven_MaxEnt07}; analysis using a predefined entropy or cross-entropy function {(with its $N^{-1}$ divisor)} can give incorrect results. Accordingly, we must extremise the logarithm of \eqref{eq:P_sys12}, subject to the constraints:
\begin{eqnarray}
\sum\limits_{i = 1}^s n_i = N_1,
\label{eq:dC0a}
\\
\sum\limits_{j = 1}^t m_j = N_2,
\label{eq:dC0b}
\\
\sum\limits_{i = 1}^s n_i \epsilon_i + \sum\limits_{j = 1}^t m_j \xi_j = E_T,
\label{eq:dC1}
\end{eqnarray}
{giving the Lagrangian:
\begin{align}
\begin{split}
L&=  \ln N_1! + \sum\limits_{i = 1}^s \bigr[ n_i \ln q_i  - \ln n_i ! \bigr] 
+  \ln N_2! + 
\\ 
& \sum\limits_{j = 1}^t \bigr[ m_j \ln r_j  - \ln m_j ! \bigr]
 - \lambda_{0a} \Bigl\{ \sum\limits_{i = 1}^s n_i - N_1 \Bigr\}  
\\ 
& - \lambda_{0b} \Bigl\{ \sum\limits_{j = 1}^t m_j - N_2 \Bigr\} - \lambda_1 \Bigl\{\sum\limits_{i = 1}^s n_i \epsilon_i + \sum\limits_{j = 1}^t m_j \xi_j - E_T \Bigr\} ,
\end{split}
\label{eq:dL1}
\end{align}} 
where $\lambda_{0a}$, $\lambda_{0b}$ and $\lambda_1$ are Lagrangian multipliers associated with constraints \eqref{eq:dC0a}-\eqref{eq:dC1}.  The extremum $\delta L=0$ of \eqref{eq:dL1} gives $\partial L / \partial n_i = 0, \forall i$ and $\partial L / \partial m_j = 0, \forall j$ for constant {$N_1$ and $N_2$, giving the inferred double distribution:}
\begin{eqnarray}
p_i^{\#} = \frac{n_i^{\#}}{N_1} = \frac{1}{N_1} \, \Lambda^{-1} \Bigl[ \frac{1}{N_1} \ln N_1! + \ln q_i - \lambda_{0a} - \lambda_1 \epsilon_i \Bigr]
\label{eq:d_phash1}
\\
\pi_j^{\#} = \frac{m_j^{\#}}{N_2} = \frac{1}{N_2} \, \Lambda^{-1} \Bigl[ \frac{1}{N_2} \ln N_2! + \ln r_j - \lambda_{0b} - \lambda_1 \xi_j \Bigr]  
\label{eq:d_phash2}
\end{eqnarray}
Again there are no factorable partition functions, hence \eqref{eq:d_phash1}-\eqref{eq:d_phash2} must be solved simultaneously with \eqref{eq:dC0a}-\eqref{eq:dC1}. The analysis is therefore constructed on a microcanonical basis (systems of particles rather than ensembles of systems), but shares some {features} of the canonical ensemble. From Sanov's \cite{Sanov_1957} theorem, each distribution \eqref{eq:d_phash1} or \eqref{eq:d_phash2} must converge to Boltzmann form (\eqref{eq:pstar} with {$\lambda_0^{KL} = \lambda_{0a}^{KL}$ or $\lambda_{0b}^{KL}$}) as $N_1 \tosp \infty$ or $N_2 \tosp \infty$.

The inferred distributions \eqref{eq:d_phash1}-\eqref{eq:d_phash2} {have important implications}.  Firstly, both share a common multiplier $\lambda_1$, hence at equilibrium, Systems 1 and 2 are of identical temperature $T=1/(k \lambda_1)$; this applies regardless of the number of particles $N_1$ and $N_2$ in each system (even for single-particle systems).  The (statistical) validity of the zeroth law of thermodynamics in small systems is therefore upheld. Secondly, each distribution $p_i^{\#}$ or $\pi_j^{\#}$ depends on the number of particles in that system, but not on the number in the other system, except {via the influence of} the common temperature.  This leads to the unsurprising conclusion that it is not necessary to impose the thermodynamic limit, or the existence of a bath containing an infinite number of particles $N_2 \tosp \infty$, for the {concept of temperature} to be valid. The {precision} of inference (e.g.\ the reproducibility of the equilibrium position) will be {less pronounced} than in the asymptotic case, due to reduced dominance of the most probable peak, but it nonetheless is meaningful. 

{Returning to the examples in Figure 3 and S1-S5, it is seen that if two non-asymptotic multinomial subsystems, with the same energy level structure and degeneracy, are of common $\lambda_1$, there must be an imbalance in the energy per particle $\langle E \rangle$ between the two subsystems. This is equivalent to the statement that the two subsystems exhibit different heat capacities ${\partial \langle E \rangle_j}/{\partial T}$, $j=1,2$. The analysis therefore reveals an apparent {paradox} in the behaviour of non-asymptotic multinomial systems at low $N$. This arises from forcing a small number of particles, with quantised energy levels, to adopt a configuration which matches the temperature of another subsystem, producing non-Boltzmann-like distributions of particles amongst energy levels and concomitant changes in the energy per particle of the subsystem.}

\section{\label{Quantum}Quantum Systems} %
For completeness, we consider the non-asymptotic form of the degenerate Maxwell-Boltzmann (MB), Bose-Einstein (BE) and Fermi-Dirac (FD) statistics of quantum physics, for the isolated system in Figure \ref{fig:systems}a.  {Although usually represented using weights 
\cite{Tolman_1938, Davidson_1962, Niven_2005, Niven_2006}, from a MaxProb perspective they are more appropriately represented using the normalised probability distributions of Brillouin \cite{Brillouin_1930, Niven_G_2009}:
\begin{eqnarray}
\mathbb{P}_{MB} = \frac{N!}{G^N} \prod\limits_{i = 1}^s {\frac{{g_i ^{n_i } }}{{n_i !}}} ,
\label{eq:P_MB} \\
\mathbb{P}_{BE} = \frac{{N!(G - 1)!}}{{(G + N - 1)!}}\prod\limits_{i = 1}^s {\frac{{(g_i  + n_i  - 1)!}}{{n_i !(g_i  - 1)!}}},
\label{eq:P_BE} \\
\mathbb{P}_{FD} = \frac{{N!(G - N)!}}{{G!}}{\rm{ }}\prod\limits_{i = 1}^s {\frac{{g_i !}}{{n_i !(g_i  - n_i )!}} }.
\label{eq:P_FD}
\end{eqnarray}
The MB statistic is identical to the multinomial \eqref{eq:multinomialP} with $q_i=g_i/G$. From the Boltzmann principle \eqref{eq:Boltzmann2}, the resulting non-asymptotic BE and FD cross-entropy functions are:
\begin{eqnarray}
\label{eq:D_BE} 
-D_{BE}  = \frac{1}{N} \sum\limits_{i = 1}^s \Bigl\{ \frac{n_i}{N} \ln \frac{N! (G-1)!}{(G+N-1)!} + \ln (g_i + n_i -1)! 
\\ \nonumber \qquad 
- \ln n_i ! - \ln (g_i -1)!   \Bigr\}
\\
\label{eq:D_FD}
-D_{FD}  = \frac{1}{N} \sum\limits_{i = 1}^s \Bigl\{ \frac{n_i}{N} \ln \frac{ N! (G-N)!}{G!} - \ln (g_i - n_i)! 
\\ \nonumber \qquad 
- \ln n_i ! - \ln g_i !   \Bigr\}
\end{eqnarray}
Maximisation of $\mathbb{P}$ or minimisation of $D$ for each case, subject to the constraints \eqref{C0}-\eqref{C1}, yields the inferred non-asymptotic distributions:
\begin{eqnarray}
n_{BE,i}^\#  &= \Lambda ^{ - 1} \Bigl[ \frac{1}{N} \ln \frac{N! (G-1)!}{(G+N-1)!} + \Lambda (g_i + n_{BE,i}^\# - 1) - \lambda _0  - \lambda _1 \epsilon _i  \Bigr]
\label{eq:n_BE} 
\\
n_{FD,i}^\#  &= \Lambda ^{ - 1} \Bigl[ \frac{1}{N} \ln \frac{ N! (G-N)!}{G!} - \Lambda (g_i - n_{FD,i} ^\#) - \lambda _0  - \lambda _1 \epsilon _i  \Bigr]
\label{eq:n_FD}
\end{eqnarray}
These can be shown to reduce to the well-known asymptotic distributions of BE and FD statistics 
\cite{Tolman_1938, Davidson_1962,  Niven_2005, Niven_2006} 
as $N \tosp \infty$.}

\begin{figure}[t]
\begin{center}
\setlength{\unitlength}{0.6pt}
  \begin{picture}(380,540)
   \put(20,0){\includegraphics[height=48mm]{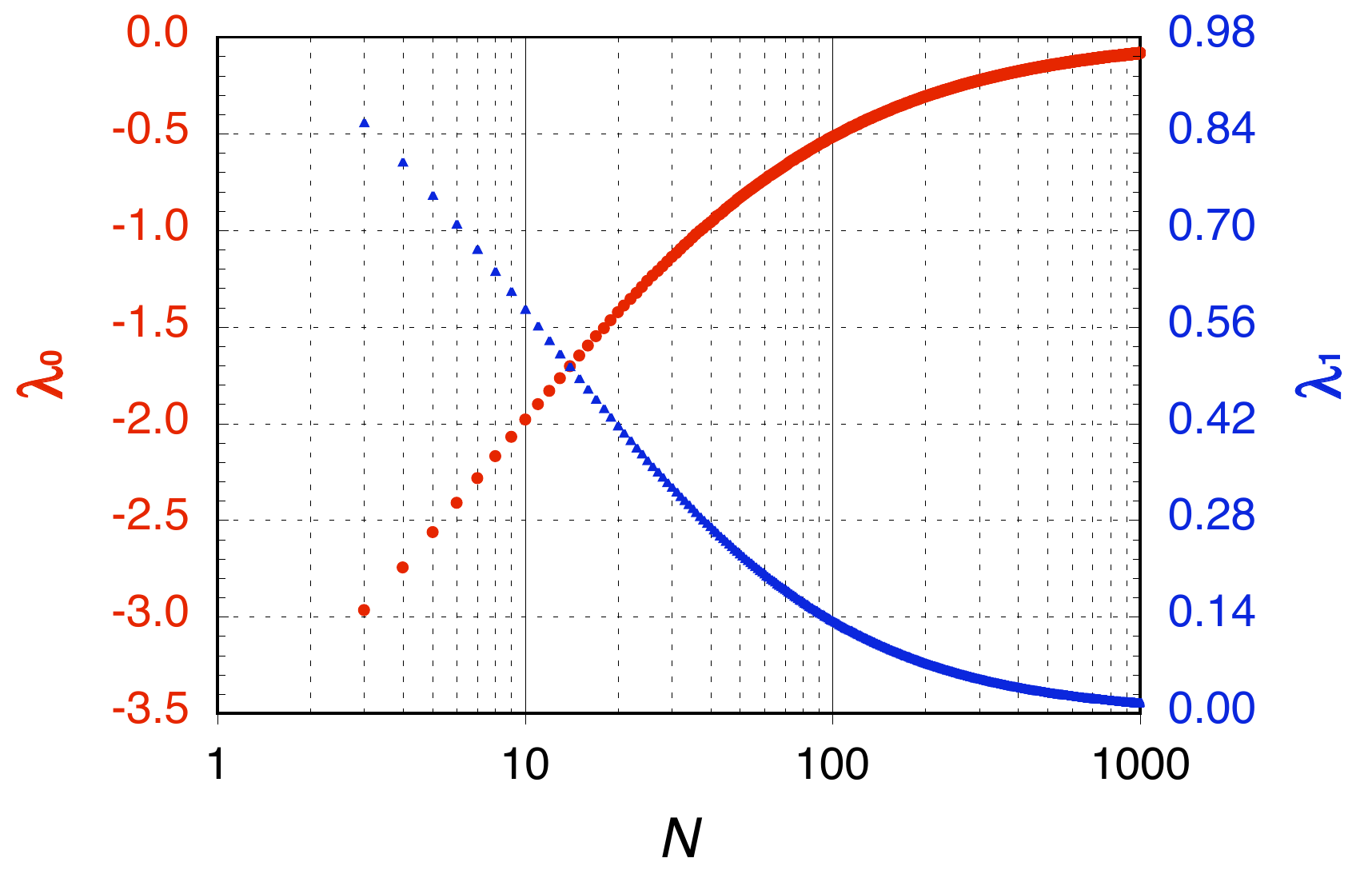} }
   \put(10,0){\small (b)}
   \put(40,280){\includegraphics[height=50mm]{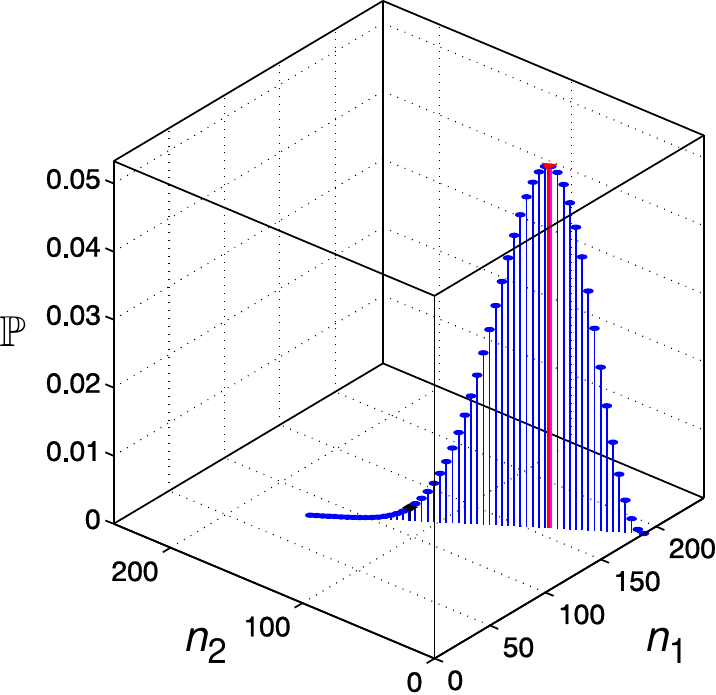} }
   \put(10,280){\small (a)}
  \end{picture}
  \end{center}
\caption{Example BE system subject to the natural and energy constraints with $\langle E \rangle=\frac{5}{3}$ (see text): (a) possible realizations for $N=243$; and (b) variation of $\lambda_0$ and $\lambda_1$ with $N$.}
\label{fig:BE}
\end{figure}

{Several plots of the probability of each realization, for the same system examined in Figures \ref{fig:multinomial}-\ref{fig:mult2} and S1-S5 but now using the BE statistic \eqref{eq:D_BE} and \eqref{eq:n_BE}, are given in Figures S6-S9.  One of these plots is shown in Figure \ref{fig:BE}a. Analysis using only the natural constraint (Figure S6) produces similar plots to the multinomial case, albeit with a broader spread of realizations. If an energy constraint is considered, the resulting plots (Figures \ref{fig:BE}a and S7-S9) display similar features to the multinomial system, in that non-asymptotic effects are extended to much higher $N$. Also, the asymptotic KL realization \eqref{eq:pstar} becomes less and less representative of the system as $N \to \infty$. However, there is one important difference to the multinomial case: the BE distribution appears to converge asymptotically to a constant ``envelope'' of possible realizations, rather than to a single, sharp peak. This implies that even in the asymptotic limit, there is considerable uncertainty in inferring the realization of the system, a result in sympathy with the} {known quantum behaviour of BE systems. Plots} {of numerical values of $\lambda_0$ and $\lambda_1$ are also given in Figures \ref{fig:BE}b and S10; these reveal consistent convergence towards asymptotic values as $N \to \infty$.}

Finally, as evident from the previous analysis (\S3), in a closed double {quantum} system (Figure \ref{fig:systems}b) the above distributions will also apply to each subsystem, with $p_i^\#$, $g_i$, $\lambda_0$ and $N$ replaced respectively by $p_i^\#$, $g_i$, $\lambda_{0a}$ and $N_1$ or $\pi_j^\#$, $g_j$, $\lambda_{0b}$ and $N_2$.  

\section{Conclusions} 
The MaxProb principle (Boltzmann's principle) is used to determine the ``most probable'' realization (macrostate) of an isolated or closed thermodynamic system, containing a small number of particles ($N \llsp \infty$), both for classical and quantum statistics. The inferred distributions provide the means to define intensive variables and construct thermodynamic relationships in systems which do not satisfy the thermodynamic limit, {using a particle-based rather than ensemble approach.} The inferred distributions will become less reproducible as $N \tosp 1$, since the most probable peak will be less and less dominant, but {still provide the best distribution for probabilistic inference. The analysis also reveals several peculiar properties of non-asymptotic systems, including a difference in the mean energy per particle between non-asymptotic subsystems in thermal equilibrium, and the asymptotic convergence of quantum (BE) systems to an ``envelope'' of possible realizations rather than a sharp peak.}

This study concerns systems with fixed $N$.  Further work is required on the non-asymptotic behaviour of systems with variable $N$ (the grand canonical ensemble).

\vspace{20pt}
\section{Acknowledgments}
The author thanks the European Commission for support as a Marie Curie Incoming International Fellow (FP6); The University of New South Wales for sabbatical leave; participants of the Facets of Entropy workshop at the University of Copenhagen, 24-26 October 2007, especially Marian Grend\'ar, Ali Ghaderi, Sergio Verdu, Filip Meysman and many others for stimulating discussions; and Bjarne Andresen for detailed comments on the manuscript.

\section*{References}

\end{document}